# Self-healing topological streams in space-time

Yudong Ren[1,2], Wenhao Li[1,2], Yuang Pan[1,2], Hongsheng Chen[1,2,*], Yihao Yang[1,2,*]

[1]College of Information Science and Electronic Engineering, Zhejiang University, Hangzhou, 310027, China

[2]ZJU-Hangzhou Global Science and Technology Innovation Center and ZJU-UIUC Institute, Zhejiang University, Hangzhou, 310027, China.

*Corresponding author. Email: yangyihao@zju.edu.cn; hansomchen@zju.edu.cn;

**Abstract**

Topological states are renowned for their robustness against perturbations. Recent advances have further introduced momentum-gap (or time) topology in time-varying media, enabling temporal topological states. However, existing studies for momentum-gap topology are largely confined to non-propagating interface states; their potential for wave transport, especially at moving boundaries, has yet to be realized. Here, we experimentally demonstrate topological streams in space-time: space-time topological edge states that propagate along a moving boundary in a dynamically modulated time-synthetic photonic lattice. These streams arise from the coexistence of energy-gap and momentum-gap topologies and break both spatial and temporal translation symmetries, conserving neither energy nor momentum. Remarkably, they reconstruct their wave profiles after strong localized spatiotemporal obstacles, enabled by an imaginary quasienergy gap and kinematic decoupling from radiative bulk channels. Our results establish a form of self-healing spatiotemporal non-Hermitian protection, showing that the temporal dimension can transform topology from pinned states and isolated events into robust, directional and self-restoring wave transport.

## Introduction

Periodicity provides one of the most fundamental organizing principles of wave and matter systems. While crystalline order is conventionally associated with structures repeated in space, recent advances have extended periodicity into time and space-time, giving rise to time crystals[1,2], photonic time crystals[3–5], and more broadly space-time photonic crystals[6]. In photonics, this shift has been particularly powerful: by modulating optical systems in time, one can engineer frequency conversion[7,8], non-reciprocal wave control[9,10], and non-equilibrium band structures[3,11,12] that are difficult to access in static media. Because time is intrinsically directional and constrained by causality, temporal periodicity can generate momentum gaps, complex-frequency spectra, and non-Hermitian dynamics[13,14] that have no direct analogue in static spatial crystals. These features open a route to topology beyond conventional energy-gap states at spatial interfaces[11,15–17]. Spatial topology supports boundary states between distinct spatial phases (Fig. 1a), whereas momentum-gap topology supports states localized at temporal interfaces[3,18,19] (Fig. 1b). When spatial and temporal periodicities coexist, energy-gap and momentum-gap topologies can combine to produce space-time-topological events[20], localized at isolated points in the space-time continuum (Fig. 1c).

However, neither purely temporal topological states nor these discrete spatiotemporal events can support continuous wave transport; they remain confined to transient moments or isolated points in space-time. Unlocking real transport requires a moving boundary in space-time that dynamically intertwines spatial and temporal topologies. In this setting, energy-gap and momentum-gap topologies are no longer separate ingredients: both space- and time-translation symmetries are continuously broken, freeing the resulting modes from the usual constraints of momentum or energy conservation. Theory predicts that such moving interfaces can host topologically protected states[21], whose non-Hermitian momentum-gap character may produce exponential growth or decay depending on the combination of spatial and temporal invariants. Yet it has remained unclear whether these highly non-equilibrium states can exist as stable, physically realizable modes and what mechanisms govern their transport.

Here, we demonstrate space-time topological edge states (STTESs) that propagate along a two-dimensional space-time boundary, forming topological streams in space-time (Fig. 1d). Realized in a dynamically modulated time-synthetic photonic lattice, these boundary modes simultaneously break spatial and temporal translation symmetries, and therefore conserve neither energy nor momentum. The coexistence of energy- and momentum-gap topologies gives the STTESs a self-healing capability. Above a kinematic threshold, their imaginary quasienergy becomes separated from the bulk continuum, so that the growth rate of the amplifying edge branch exceeds the maximum bulk gain. The resulting STTESs recover from localized defects and remain robust against strong static spatial and dynamic spatiotemporal disorder, up to a full $2\pi$ phase range. Our findings establish a novel paradigm of spatiotemporal non-Hermitian protection, opening promising avenues for resilient wave manipulation in highly non-equilibrium time-varying media.

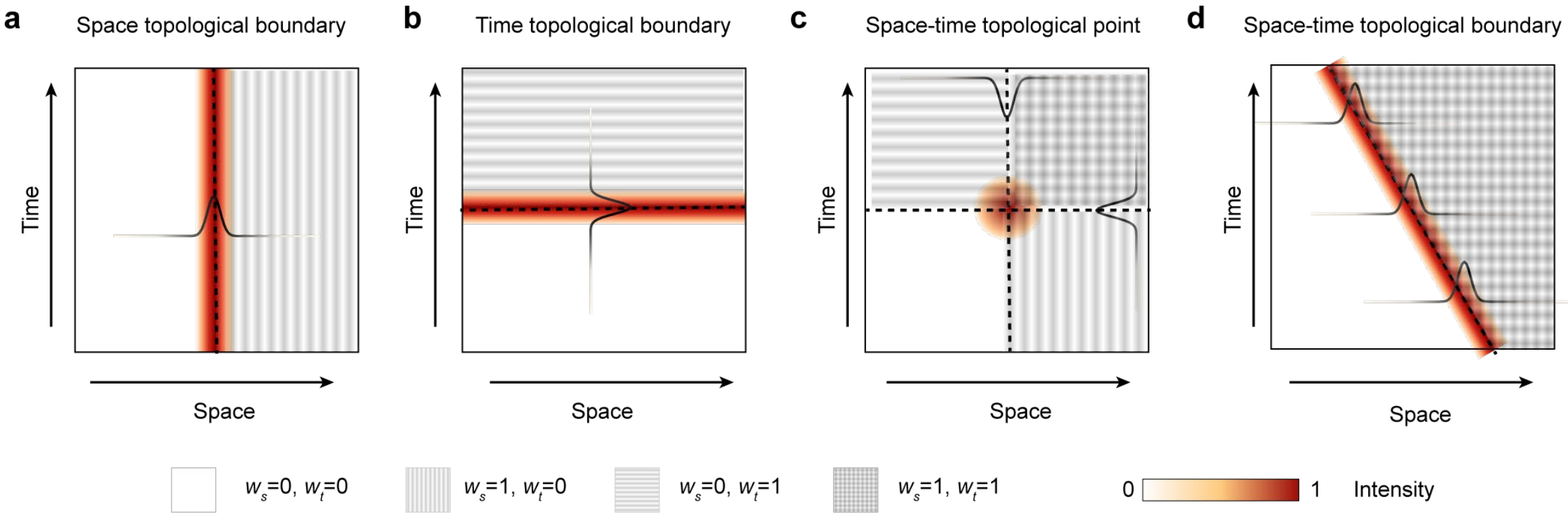

**Fig. 1. Evolution of topological states across the space-time landscape.** (**a**, **b**) One-dimensional spatial and temporal topological states occur at interfaces where the spatial and temporal invariants, respectively, change. (**c**) A zero-dimensional space-time topological event occurs at the intersection of the two interfaces. (**d**) Space-time topological edge states propagate along a moving boundary between bulk phases with distinct space-time topology.

## Space-time topological photonic lattice

Our experiments are carried out in a space-time photonic lattice implemented via coupled fiber loops[22]. Operating at the telecommunication wavelength of 1550 nm, two fiber loops of slightly different lengths are connected by a variable beamsplitter (VBS), as illustrated in Fig. 2a (see Fig. S1 in the Supplementary Information for the comprehensive experimental layout). Optical pulses circulating in the loops experience splitting, delay, and recombination, mapping onto the dynamics of a one-dimensional synthetic lattice (see Supplementary Information). The shorter and longer loops correspond to two internal pseudo-spin states, denoted as $|U\rangle$ (left-moving) and $|V\rangle$ (right-moving), respectively. The discrete position $x$ in the lattice is encoded in the time delay between pulses, while the round-trip number $t$ serves as the discrete time step. The evolution of a single pulse follows the discrete-time quantum walk described by

$$|\psi(x,t+1)\rangle = U_c\,|\psi(x,t)\rangle, \tag{1}$$

where the $|\psi(x,t)\rangle = \sum_x(u_x^t|n\rangle\otimes|U\rangle + v_x^t|n\rangle\otimes|V\rangle)$ is the wavefunction, with $u_x^t$ ($v_x^t$) denoting the pulse amplitude at lattice position $x$ and time step $t$. The one-step time-evolution operator $U_c$ can be factorized as

$$U_c = SG(g)C(\beta), \tag{2}$$

where $S = \sum_x(|x-1\rangle\langle x|\otimes|U\rangle\langle U|+|x+1\rangle\langle x|\otimes|V\rangle\langle V|)$ is the conditional translation operator,

$$C(\beta) = I_x \otimes \begin{pmatrix} \cos\beta & i\sin\beta \\ i\sin\beta & \cos\beta \end{pmatrix},\ G(g) = I_x \otimes \begin{pmatrix} e^g & 0 \\ 0 & e^{-g} \end{pmatrix},\ I_x = \sum_x |x\rangle\langle x|,$$

is the position-dependent coin operator, non-Hermitian modulation operator and identity, respectively. The parameter $\beta$ controls the splitting ratio of the variable beamsplitter, while $g$ represents the non-Hermitian gain-loss parameter. By dynamically modulating $\beta$ and $g$, we can independently control the energy-gap topology (space topology) and momentum-gap topology (time topology) of the synthetic system[20,23]. To establish the spatial topology, we define $\Delta\beta$ as the spatial modulation amplitude or dimerization strength. Specifically, alternating beamsplitter ratio $\beta$ between $\beta_1 = \pi/4 + \Delta\beta/2$ and $\beta_2 = \pi/4 - \Delta\beta/2$ in consecutive time steps implements a Floquet Su-Schrieffer-Heeger (SSH) model (Fig. 2b), thereby realizing a spatial topological lattice. To introduce time topology, we apply a four-step gain-loss protocol (see Fig. 2c), which opens distinct momentum gaps accompanied by non-trivial momentum-gap topology. The simultaneous combination of both spatial and temporal modulations yields a full space-time topology, supporting the propagating space-time topological streams investigated in this work. By solving the eigenvalue equation of the time-evolution operator in momentum space, we obtain the quasienergy bands, which exhibit simultaneous energy and momentum gaps (see Supplementary Information). Note that this space-time topology can also be successfully realized using a simplified two-step protocol (see Supplementary Information).

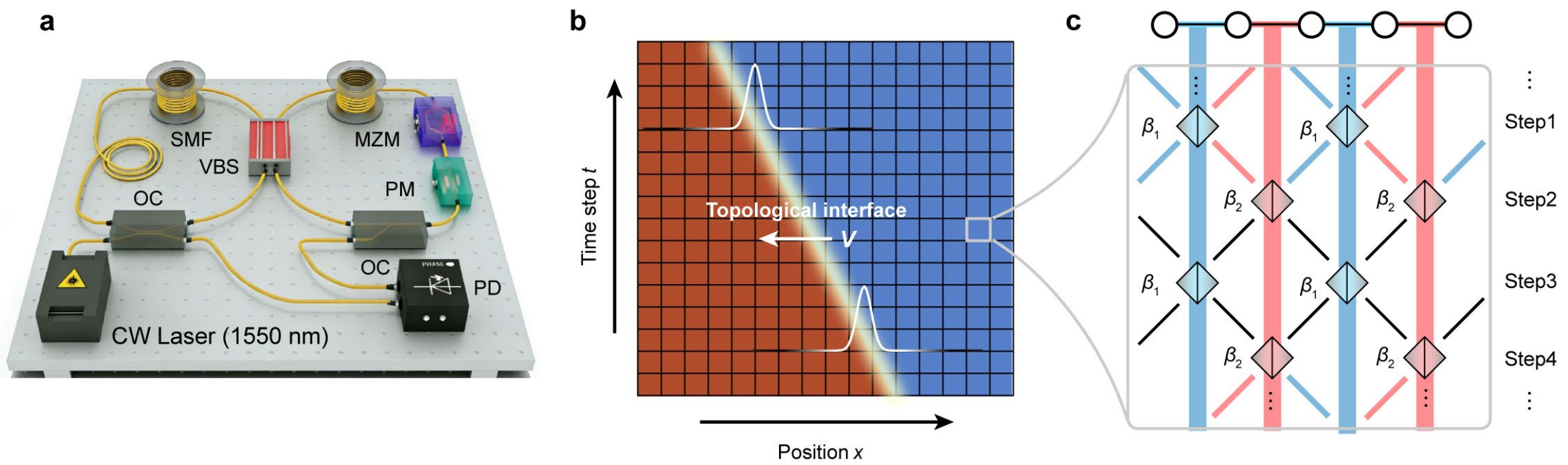

**Fig. 2. Experimental platform and dynamic modulation configurations.** (**a**) Schematic of the time-synthetic photonic lattice based on coupled fiber loops. The system consists of two fiber loops with slightly different lengths, where PM, MZM, OC, VBS, SMF, and PD denote the phase modulator, Mach-Zehnder modulator, optical coupler, variable beamsplitter, single-mode fiber, and photodetector, respectively. (**b**) Illustration of the moving topological boundary. The red and blue regions on each side represent two bulks with distinct space-time topologies. The boundary moves at a velocity $v$, breaking both spatial and temporal translation symmetries. (**c**) Detailed four-step protocol for realizing space-time topology within each space-time unit cell.

**Topological space-time edge states**

While purely spatial and temporal interfaces have been individually demonstrated in prior studies, here we investigate a more generalized moving interface that separates the two-dimensional space-time continuum into two half-infinite domains. We focus on a moving boundary characterized by a uniform velocity of $v$ = -0.5, as illustrated in Fig. 2b. The corresponding bulk bands, obtained by solving the eigenvalue problem of Eq. (2), exhibit simultaneous energy and momentum gaps (Fig. 3a). By independently choosing the spatial and temporal invariants of the two bulk domains, one can construct fully space-time-topological, spatial-topological but temporal-trivial, and spatial-trivial but temporal-topological interfaces. These configurations exhibit distinct gain-loss characteristics. Here, we focus on the configurations supporting the space-time topological edge states (STTESs). We calculate their dispersion using a space-time supercell (Fig. 3a, b). The energy-gap topology provides transverse confinement of the interface mode, associated with an imaginary transverse momentum, whereas the temporal gain-loss modulation renders its quasienergy complex. The real part of the STTES dispersion is linear, with $\partial_k \mathrm{Re}[E] = -v$, indicating that the group velocity of every supported edge branch is locked to the velocity of the moving interface. The imaginary part of the dispersion exhibits a gain-loss duality (Fig. 3b), which originates from the underlying symmetry of the space-time moving boundary (see Supplementary Information).

In the absence of temporal topology ($g = 0$), the edge states possess purely real quasienergies, and the corresponding wavefunctions oscillate periodically in space and time without any amplification or decay. Activating the temporal gain-loss protocol ($g \neq 0$) produces complex quasienergies. The real edge-state dispersion crosses both the energy and momentum gaps of the bulk spectrum (Fig. 3a), implying dual topological protection. Its real dispersion fixes the propagation trajectory, whereas its imaginary quasienergy governs the exponential amplification or attenuation along that trajectory. Because the moving boundary breaks both spatial and temporal translation symmetries, the mode can exchange both momentum and energy with the modulation and is not subject to the conventional conservation laws of a static lattice.

A particularly remarkable feature arising from this spatiotemporal framework is that these edge modes can exhibit exceptional amplification (or decay) rates that vastly exceed the maximum imaginary eigenvalues of the surrounding bulk continuum (Fig. 3b). This spectral separation is absent for the spatial-topological but temporal-trivial control case, whose edge-state gain remains bounded by the bulk spectrum. To quantify the gain isolation, we map out the phase diagram of the imaginary gap $\Delta\gamma$ as a function of the modulation parameters $\Delta\beta$ and $g$ (Fig. 3c), revealing that a prominent imaginary gap opens when $\Delta\beta$ exceeds a critical threshold of $\Delta\beta_c \approx \pi/6$ (see Supplementary Information). This threshold behavior can be captured by an effective kinematic framework: in the co-moving frame of the interface, the bulk dispersion undergoes a geometric tilting due to the Doppler shift $-vk$. The STTES becomes gain-isolated when the Doppler-tilted bulk dispersion no longer contains a mode whose group velocity matches that of the moving interface. This kinematic mismatch suppresses radiative leakage from the edge state into the bulk continuum. The edge mode consequently crosses over from a leaky interface state to a gain-isolated amplified bound state, which provides the spectral basis for its self-healing behavior. The corresponding spatiotemporal field distributions for both the amplifying and decaying configurations are visualized in Fig. 3d and e, respectively.

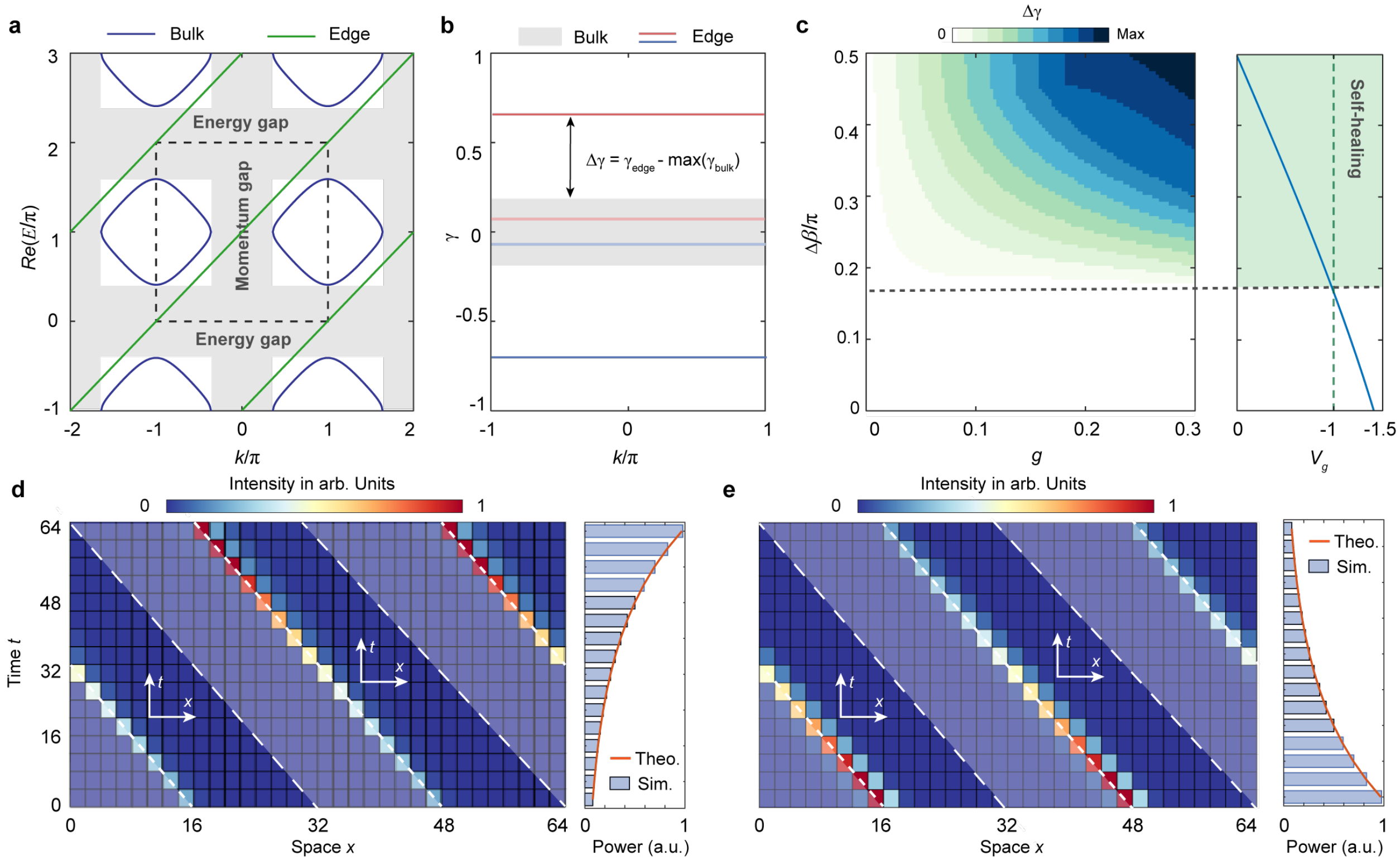

**Fig. 3. Space-time topological streams.** (**a**) Dispersion of bulk and the STTESs. The blue curves represent the bulk band structure of the spatiotemporal lattice with reconfigurable spatial and temporal topologies, while the gray-shaded regions represent the energy and momentum gaps. The green lines represent the dispersion of the STTESs, which simultaneously cross the energy and momentum gaps, demonstrating dual topological protection. (**b**) Imaginary part of dispersion of the STTESs. The gray shading indicates the continuum of bulk gain and loss, while the red and blue curves correspond to the positive (gain) and negative (loss) imaginary quasienergy branches of the edge states, respectively. (**c**) Imaginary gap $\Delta\gamma$ as a function of the dimerization strength $\Delta\beta$ and non-Hermitian parameter $g$. A positive gap opens above the kinematic threshold $\Delta\beta_c \approx \pi/6$. (**d** and **e**) Calculated spatiotemporal profiles of the amplifying (d) and decaying (e) STTESs at $Re(E) = 0$ and $k = 0$. For each configuration, the left panel illustrates the field distribution in the space-time plane, and the right panel plots the corresponding intensity evolution across discrete temporal steps.

## Self-healing property of space-time topological streams

The autonomous shape restoration of a scattered wave packet, known as self-healing (Fig. 4a), has been widely documented in structured waves, such as diffraction-free Bessel or Airy beams[24–30]. However, these extended waves rely on transverse energy flow from their unscattered, infinite-energy side lobes to reconstruct the main profile. In contrast, a fundamental physical constraint dictates that strictly localized, normalizable bound states are generally unable to perfectly self-heal within conventional Hermitian regimes, as scattering into radiation channels leads to irreversible energy leakage. Non-Hermitian dynamics offer a powerful mechanism to circumvent this restriction by utilizing gain-guided attractors[31]. Here, we experimentally demonstrate the anomalous self-healing property of our STTES within a time-synthetic photonic lattice. To rigorously quantify the spatiotemporal self-healing dynamics, we define $|\varphi(t)\rangle$ to be the evolved wave function in the absence of any defect, and $|\psi(t)\rangle$ as the perturbed wavefunction in the presence of a space-time obstacle. In a general (non-self-healing) case, $|\psi(t)\rangle$ can largely deviate forever from the unperturbed solution $|\varphi\rangle$ after interaction with the space-time obstacle. Conversely, a topological wave packet is designated as self-healing if the spatiotemporal deviation $|\zeta(t)\rangle \equiv |\psi(t)\rangle - |\varphi(t)\rangle$ is asymptotically much smaller than $|\varphi(t)\rangle$ as $t \to \infty$ for a broad class of localized space-time obstacles. This is equivalent to requiring $\lim_{t\to\infty} \varepsilon(t) = 0$, where

$$\epsilon(t) = \sum_x | I_s(x,t) - I_r(x,t) |, \tag{3}$$

where $I_s(x,t) = |\psi(x,t)|^2$ and $I_r(x,t) = |\varphi(x,t)|^2$ are the normalized intensities of the perturbed and unperturbed wave packets, respectively. To systematically characterize the self-healing property, we map out a phase diagram in the parameter space of $\Delta\beta$ and $g$ (Fig. 4b), where the color scale indicates the self-healing rate $R = 1-\varepsilon(t_f)/\varepsilon_{\max}$ (with $t_f$ sufficiently large). This result is in excellent agreement with Fig. 3c, demonstrating that self-healing occurs when $\Delta\beta$ exceeds a critical value $\Delta\beta_c$.

To experimentally study the self-healing property of the space-time topological streams, we excite the photonic quantum walk at a single site, i.e., $\psi(x,0) = \delta_{x,40}$, localized right at the topological boundary. As expected, a space-time topological stream forms at the interface, propagating along the boundary with a constant velocity of $v = -0.5$. In the absence of any obstacle, the wave packet propagates indefinitely while strictly maintaining its spatial profile. Here, we evaluate and compare two distinct configurations: the STTES possessing full space-time topology, and a control case featuring space-only topology (see the detailed parameters in Supplementary Information). To demonstrate this resilience, we introduce a composite space-time obstacle applied within the temporal window $18 < t < 21$ and in the spatial region $29 < x < 36$ (Fig. 4c). This obstacle simultaneously incorporates three distinct types of disorder: an absorbing potential, a perturbed coupling ratio, and random phase fluctuation (see Supplementary Information). First, we examine the control case of the STTES with space-only topology, which lacks full spatiotemporal protection. Figure 4c displays the measured temporal evolution of the normalized wave function for this configuration. Upon encountering the space-time obstacle, the wave packet undergoes severe, irreversible distortion and fails to recover its original profile. Instead, the majority of the optical energy scatters into the bulk modes on the right side of the boundary. Correspondingly, the deviation $\varepsilon(t)$ rises sharply upon impact and remains pinned at a high finite value thereafter (Fig. 4c, right), indicating the absence of self-healing.

In stark contrast, when the full space-time topology is present, the STTES exhibits robust self-healing. As shown in Fig. 4d, although the wave packet is initially deformed and fragmented within the obstacle zone, it gradually reassembles and completely restores its original shape over several subsequent evolutionary steps. The deviation $\varepsilon(t)$ decays to zero post-interaction, confirming a successful and clean recovery (additional assessments against various obstacle profiles are detailed in the Supplementary Information). This self-healing mechanism is governed by the spectral hierarchy of the non-Hermitian Floquet operator. Because the STTES possesses an imaginary quasienergy that strictly exceeds the upper bound of the bulk spectrum across the entire Brillouin zone, the system undergoes a process of non-Hermitian spectral purification upon scattering: any energy diverted into bulk modes experiences rapid modal attenuation due to the eigenvalue hierarchy. Consequently, the STTES acts as a potent dynamical attractor. Regardless of the localized profile or timing of the spatiotemporal defect, the system asymptotically collapses back onto this dominant Floquet eigenmode, successfully reconstructing its complete wave profile so long as the topological boundary remains within the causal light cone.

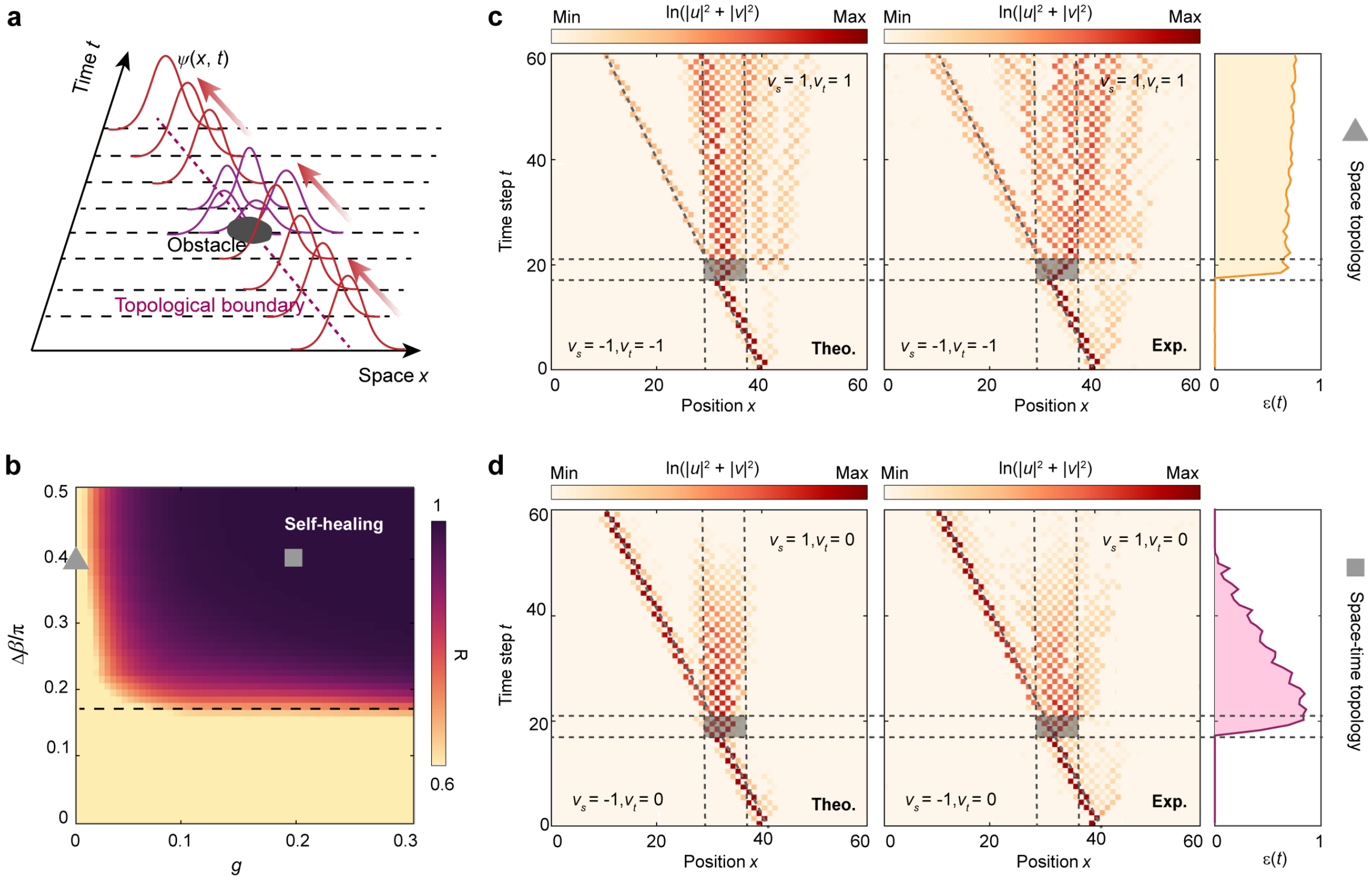

**Fig. 4. Self-healing dynamics of space-time topological streams.** (**a**) Concept diagram for the wavefunction of the moving space-time topological states and their self-healing property. The space-time obstacle has a limited length and time, and the dashed line indicates the moving topological boundary. After a characteristic healing time, the wave packet reconstructs its original shape. (**b**) Phase diagram of self-healing. The color scale represents the self-healing rate $R$. (**c**) Temporal evolution of the normalized wavefunction of space-time topological edge states with only space topology ($\Delta\beta$ = 0.4π, $g$ = 0). The left panels display a side-by-side comparison between the theoretical (Theo.) and experimental (Exp.) intensity evolutions; the right panel shows the corresponding temporal evolution of the function $\varepsilon(t)$, defined by Eq. (3), which measures the deviation of the evolved wave function from the unperturbed space-time topological edge states. (**d**) Same as (c) but for edge states with space-time topology ($\Delta\beta$ = 0.4π, $g$ = 0.2). The wave packet recovers after the obstacle, and $\varepsilon(t)$ decays to zero.

## Robustness of space-time topological streams

The inherent robustness of topological states against disorder is their most celebrated feature, universally demonstrated across diverse platforms ranging from quantum electronic systems to classical photonic and phononic devices. However, in conventional topological systems, strong spatial or temporal randomness frequently triggers the catastrophic breakdown of this protection via Anderson localization or backscattering, severely degrading topological transport. For instance, in time-reversal-symmetric systems such as valley-Hall photonic crystals, severe scattering out of the topological waveguides inevitably occurs under real-world disorder conditions[32]. Even for Chern modes with broken time-reversal symmetry, this protection remains strictly bounded by the intrinsic bandgap size. This limitation is fundamentally rooted in gap-closing dynamics: while weak disorder leaves the energy gap intact[33], increasing disorder strengths narrow and ultimately close the gap, causing the topological state to dissolve into Anderson-localized configurations. This bottleneck is equally pronounced in non-Hermitian point-gap topology, where strong disorder nullifies the winding number and causes the non-Hermitian skin modes to vanish[34], and in previously reported space-time crystals where gap-closing induces the collapse of topological events[20]. In stark contrast, the space-time topological streams investigated here exhibit an extraordinary resilience, remaining robust against both space and space-time disorder even at strengths that far exceed the conventional bandgap limit, as demonstrated below.

Experimentally, we introduce a uniformly distributed random phase disorder $\phi_u \in [-W/2, W/2)$ into the left-moving channel $u_x^t$ either purely in space $x$ (space disorder) or across both space and time ($x$, $t$) (space-time disorder), to explore their interplay with the edge dynamics. First, we examine the impact of space disorder. As shown in Fig. 5a, under strong spatial randomness, the control edge state protected by space-only topology is severely disrupted; Anderson localization dominates the dynamics as the energy gap closes, triggering a topological phase transition. In contrast, the STTES endowed with full space-time topology exhibits extraordinary robustness against disorder, demonstrating robust transport without noticeable time-reflections scattered into the bulk of the space-time lattice (Fig. 5a). The right panel of Fig. 5a plots the corresponding temporal evolution of the structural deviation $\varepsilon(t)$, clearly illustrating that the space-time topology provides a significantly enhanced resistance to scattering compared to its space-only counterpart.

Remarkably, the STTES is not only immune to static space disorder but also exhibits exceptional resilience against dynamic space-time disorder. Unlike space disorder, which triggers conventional Anderson localization, space-time disorder (or dynamic noise) fundamentally disrupts the phase coherence along distinct interference pathways. Because spatiotemporal randomness breaks both energy and momentum conservation, it acts as a severe scattering mechanism that typically knocks waves out of the bandgap and into the bulk bands. Upon introducing spatiotemporal disorder, the energy initially localized within the space-only topological boundary continuously bleeds into the bulk modes of the space-time lattice (Fig. 5c). However, the STTESs protected by space-time topology do not exhibit any noticeable time-reflections that scatter energy into the bulk of the space-time lattice (Fig. 5c), highlighting the space-time topological protection. To rigorously verify the statistical universality of this resilience, we then numerically extract the time-averaged deviation for the edge modes for 1,000 realizations of disorder, and plot its magnitude versus $W$, as shown in Fig. 5b (space disorder) and d (space-time disorder). In the clean limit ($W = 0$), the edge states exist in both cases and are unperturbed. However, a stark divergence emerges at the maximum disorder limit ($W = 2\pi$), where the phase fluctuations span the entire quasienergy spectrum, a scale far exceeding the intrinsic topological bandgap. While the space-only topological edge states completely break down under this extreme randomness, the STTES remains exceptionally robust and well-preserved, firmly validating the power of non-Hermitian space-time topological protection in highly imperfect environments.

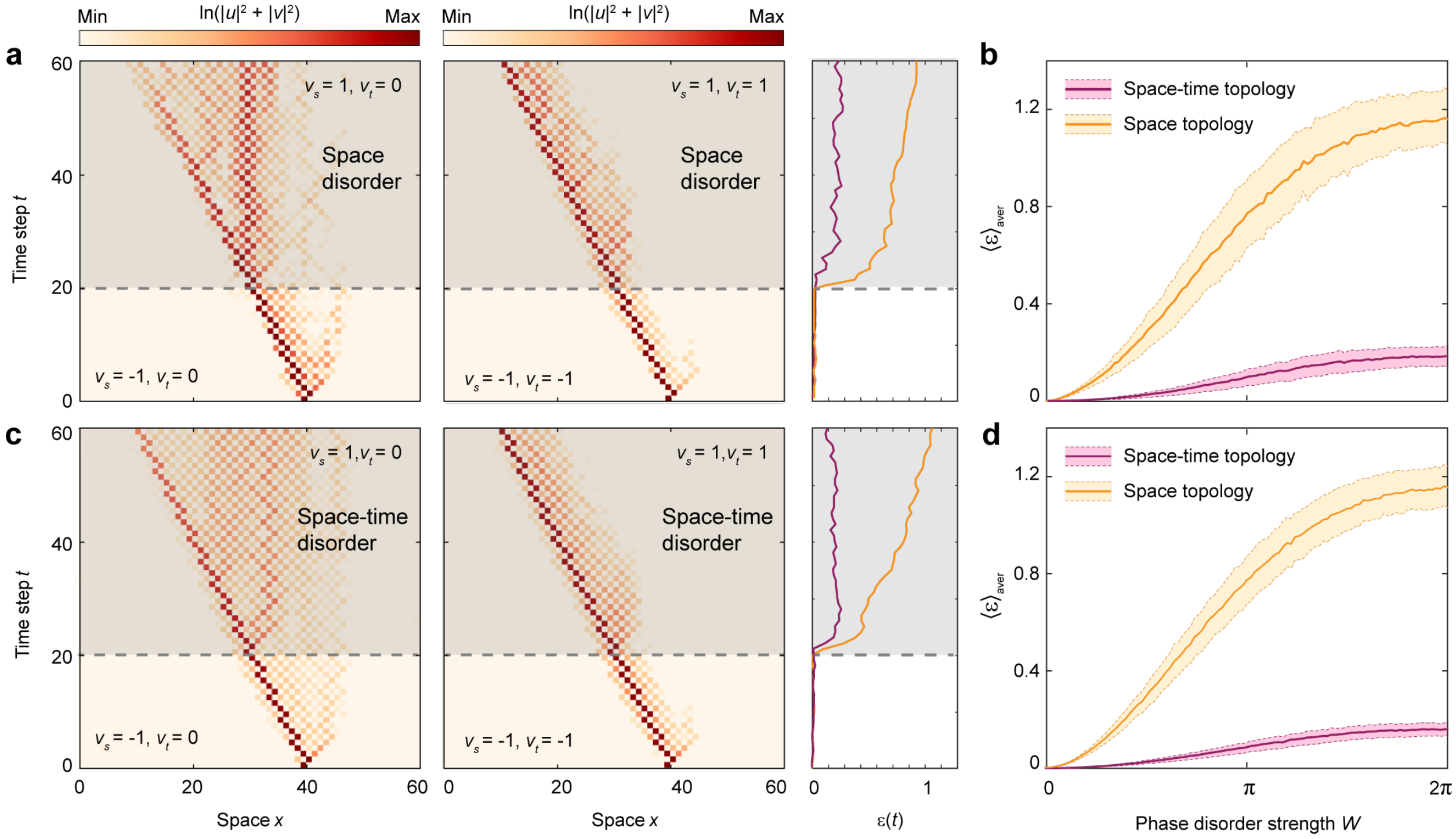

**Fig. 5. Transport robustness of space-time topological streams.** (**a**) Dynamical behavior under severe spatial disorder initiated at $t = 20$. Left: the control edge state with space-only topology undergoes breakdown, where the edge channel is disrupted and Anderson localization dominates the dynamics. Middle panel: under the identical disorder profile, the STTES with full space-time topology remains exceptionally robust. Right panel: corresponding temporal evolution of the structural deviation $\varepsilon(t)$ for both configurations. (**b**) Time-averaged deviation $\langle\varepsilon\rangle$ (over $t = 51$ to 60) as a function of spatial disorder strength $W$, averaged over 1,000 disorder realizations (solid lines). The dashed lines indicate the first and third quartiles ($Q_1$ and $Q_3$). Edge states with space-time topology exhibit low deviation and survive disorder strengths up to a full 2π. (**c**) Same as (a), but for space-time disorder. (**d**) Same as (b), but for space-time disorder.

## Conclusion

In conclusion, we experimentally demonstrate self-healing space-time topological edge states residing on moving interfaces within a time-synthetic photonic lattice. While conventional topological platforms fundamentally rely on the conservation of either energy or momentum, these spatiotemporal edge states break both space- and time-translation symmetries, yet exhibit unprecedented resilience against severe disorders and localized defects. Remarkably, although the system features only a single spatial dimension, a regime that conventionally precludes robust topological edge transport, the synergy of space-time modulation and non-Hermitian engineering enables unhindered, directional wave propagation along the moving boundary. Driven by a distinct topological gain margin over the bulk spectrum, these states possess an intrinsic capability for autonomous self-healing and topological immunity against extreme spatial and spatiotemporal randomness. More broadly, the paradigm established here can be seamlessly extended to higher-dimensional moving boundaries separating domains with distinct space-time topologies, and these exotic dispersion relations may be mapped onto various time-varying or Floquet platforms. Our findings unlock a new frontier for resilient wave manipulation in non-equilibrium media, paving the way for advanced applications in temporal cloaking, fault-tolerant optical signal routing, and non-reciprocal dynamic wave engineering.

**Acknowledgments**

We thank Chengwei Qiu for his insightful discussions. This work was sponsored by the Key Research and Development Program of the Ministry of Science and Technology under Grant Nos. 2022YFA1405200 (Y.Y.), 2022YFA1404900 (Y.Y.), 2022YFA1404704 (H.C.), and 2022YFA1404902 (H.C.); the National Natural Science Foundation of China (NNSFC) under Grant Nos. 61975176 (H.C.) and U25D8017 (Y.Y.); the Key Research and Development Program of Zhejiang Province under Grant No. 2022C01036 (H.C.); the Fundamental Research Funds for the Central Universities under Grant No. 2021FZZX001-19 (Y.Y.); the Fundamental Research Funds for the Zhejiang Provincial Universities under Grant No. 226-2025-00231 (Y.Y.); and the Science Challenge Project under Grant No. TZ2025015 (Y.Y.).

**Author contributions**

Y.Y. and Y.R. conceived the study and initiated the project. Y.R. developed the theoretical framework, designed the experimental setup, and performed the experiments with assistance from W.L. and Y.P. Y.R. conducted the numerical simulations and carried out the data analysis. The manuscript was drafted by Y.R. and revised by Y.Y., H.C. The research was supervised by Y.Y., H.C. All authors discussed the results, reviewed the manuscript, and approved the final version.

**Competing interests:** Authors declare that they have no competing interests.

**Data and materials availability:** All data are available in the main text or the supplementary materials.